\documentclass[11pt]{article}

\usepackage{graphicx}
\usepackage{amssymb}
\usepackage[english]{babel}
\usepackage[leqno]{amsmath}
\usepackage{xspace}
\usepackage{hyperref}
\hypersetup{colorlinks=true,allcolors=black}
\usepackage{xcolor}
\usepackage{lineno}
\begin{document}

\centerline{\LARGE EUROPEAN ORGANIZATION FOR NUCLEAR RESEARCH}

\vspace{10mm} {\flushright{
CERN-EP-2023-245 \\
October 30, 2023\\
Revised February 9, 2024\\
}}

\vspace{40mm}

\begin{center}

\boldmath

{\bf {\Large\boldmath{First observation and study of the $K^{\pm} \rightarrow  \pi^{0} \pi^{0} \mu^{\pm} \nu$ decay}}}

\unboldmath

\end{center}

\begin{center}

{\Large The NA48/2 Collaboration
\footnote{Corresponding authors: A.~Korotkova, D.~Madigozhin, \\
email : anna.korotkova@cern.ch, dmitry.madigozhin@cern.ch }}\\

\end{center}

\begin{abstract}
The NA48/2 experiment at CERN reports the first observation of the $K^{\pm} \rightarrow  \pi^{0} \pi^{0} \mu^{\pm} \nu$ decay  
based on a sample of 2437 candidates with 15\% background contamination  collected in 2003--2004. The decay branching 
ratio in the kinematic region of the squared dilepton mass above 
$0.03$~GeV$^2/c^4$ is measured to be 
$(0.65 \pm 0.03) \times 10^{-6}$. The extrapolation to the full kinematic space, using a specific model, 
is found to be 
$(3.45 \pm 0.16) \times 10^{-6}$, in agreement with chiral perturbation theory predictions.
\end{abstract}

\begin{center}

{\it Accepted for publication in JHEP.}

\end{center}

\newpage

\newcommand{\orcimg}{\raisebox{-0.3\height}{\includegraphics[height=\fontcharht\font`A]{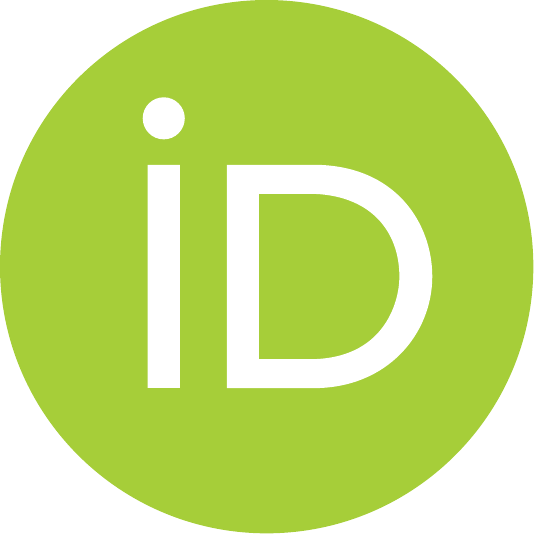}}}
\newcommand{\orcid}[1]{\href{https://orcid.org/#1}{\orcimg}}

\centerline{\bf The NA48/2 Collaboration} 
\vspace{1.5cm}
%
%

\begin{raggedright}
\noindent
{\bf Cavendish Laboratory, University of Cambridge, Cambridge, United Kingdom}\\
 J.R.~Batley\orcid{00000-0001-7658-7766},
 G.~Kalmus,
 C.~Lazzeroni$\,${\footnotemark[1]}\orcid{0000-0003-4074-4787},
 D.J.~Munday$\,${\footnotemark[1]},
 M.W.~Slater$\,${\footnotemark[1]}\orcid{0000-0002-2687-1950},
 S.A.~Wotton\orcid{0000-0003-4543-8121}
\vspace{0.5cm}

{\bf CERN, Gen\`eve, Switzerland}\\
 R.~Arcidiacono$\,${\footnotemark[2]}\orcid{0000-0001-5904-142X},
 G.~Bocquet,
 N.~Cabibbo$\,$\renewcommand{\thefootnote}{\fnsymbol{footnote}}\footnotemark[2]\renewcommand{\thefootnote}{\arabic{footnote}},
 A.~Ceccucci\orcid{0000-0002-9506-866X},
 D.~Cundy$\,${\footnotemark[3]}, 
 V.~Falaleev$\,${\footnotemark[4]}\orcid{0000-0003-3150-2196},
 M.~Fidecaro$\,$\renewcommand{\thefootnote}{\fnsymbol{footnote}}\footnotemark[2]\renewcommand{\thefootnote}{\arabic{footnote}}\orcid{0000-0002-4479-8689},
 L.~Gatignon$\,${\footnotemark[5]}\orcid{0000-0001-6439-2945},
 A.~Gonidec,
 W.~Kubischta, 
 A.~Maier,
 A.~Norton$\,${\footnotemark[6]}\orcid{0000-0001-5959-5879},
 M.~Patel$\,${\footnotemark[7]},
 A.~Peters
\vspace{0.5cm}

{\bf The Enrico Fermi Institute, The University of Chicago, Chicago, USA}\\
 E.~Monnier$\,${\footnotemark[8]}\orcid{0000-0002-2551-5751},
 E.~Swallow$\,$\renewcommand{\thefootnote}{\fnsymbol{footnote}}\footnotemark[2]\renewcommand{\thefootnote}{\arabic{footnote}},
 R.~Winston$\,${\footnotemark[9]}
\vspace{0.5cm}

{\bf Department of Physics and Astronomy, University of Edinburgh, Edimburgh, United Kingdom}\\
 P.~Rubin$\,${\footnotemark[10]}\orcid{0000-0001-6678-4985},
 A.~Walker
\vspace{0.5cm}

{\bf Dipartimento di Fisica e Scienze della Terra dell'Universit\`a e Sezione dell'INFN di Ferrara, Ferrara, Italy}\\
 P.~Dalpiaz,
 C.~Damiani,
 M.~Fiorini\orcid{0000-0001-6559-2084},
 M.~Martini,
 F.~Petrucci\orcid{0000-0002-7220-6919}, 
 M.~Savri\'e\orcid{0000-0002-8214-9671},
 M.~Scarpa,
 H.~Wahl$\,${\footnotemark[11]}\orcid{0000-0003-0354-2465}
\vspace{0.5cm}

{\bf Sezione dell'INFN di Ferrara, Ferrara, Italy}\\
 W.~Baldini\orcid{0000-0001-7658-8777},
 A.~Cotta Ramusino\orcid{0000-0003-1727-2478},
 A.~Gianoli\orcid{0000-0002-2456-8667}
\vspace{0.5cm}

{\bf Dipartimento di Fisica, Universit\`a di Firenze e Sezione dell'INFN di Firenze, Sesto Fiorentino, Italy}\\
 M.~Calvetti,
 E.~Celeghini,
 E.~Iacopini\orcid{0000-0002-5605-2497},
 M.~Lenti\orcid{0000-0002-2765-3955},
 G.~Ruggiero\orcid{0000-0001-6605-4739}
\vspace{0.5cm}

{\bf Sezione dell'INFN di Firenze, Sesto Fiorentino, Italy}\\
 A.~Bizzeti$\,${\footnotemark[12]}\orcid{0000-0001-5729-5530},
 M.~Veltri$\,${\footnotemark[13]}\orcid{0000-0001-7917-9661}
\vspace{0.5cm}

{\bf Institut f\"ur Physik, Universit\"at Mainz, Mainz, Germany}\\
 M.~Behler,
 K.~Eppard,
 M.~Hita-Hochgesand,
 K.~Kleinknecht,
 P.~Marouelli, 
 L.~Masetti\orcid{0000-0002-0038-5372},
 U.~Moosbrugger,
 C.~Morales Morales,
 B.~Renk,
 M.~Wache, 
 R.~Wanke\orcid{0000-0002-3636-360X},
 A.~Winhart
\vspace{0.5cm}

{\bf Department of Physics and Astronomy, Northwestern University, Evanston, USA}\\
 D.~Coward$\,${\footnotemark[14]}\orcid{0000-0001-7588-1779},
 A.~Dabrowski$\,${\footnotemark[15]}\orcid{0000-0003-2570-9676},
 T.~Fonseca Martin,
 M.~Shieh,
 M.~Szleper$\,${\footnotemark[16]}\orcid{0000-0002-1697-004X}, 
 M.~Velasco\orcid{0000-0002-1619-3121},
 M.D.~Wood$\,${\footnotemark[14]}
\vspace{0.5cm}

{\bf Dipartimento di Fisica dell'Universit\`a e Sezione dell'INFN di Perugia, Perugia, Italy}\\
 G.~Anzivino\orcid{0000-0002-5967-0952},
 E.~Imbergamo,
 A.~Nappi$\,$\renewcommand{\thefootnote}{\fnsymbol{footnote}}\footnotemark[2]\renewcommand{\thefootnote}{\arabic{footnote}},
 M.~Piccini\orcid{0000-0001-8659-4409},
 M.~Raggi$\,${\footnotemark[17]}\orcid{0000-0002-7448-9481},
 M.~Valdata-Nappi
\vspace{0.5cm}

{\bf Sezione dell'INFN di Perugia, Perugia, Italy}\\
 P.~Cenci\orcid{0000-0001-6149-2676},
 M.~Pepe\orcid{0000-0001-5624-4010},
 M.C.~Petrucci
\vspace{0.5cm}

{\bf Dipartimento di Fisica dell'Universit\`a e Sezione dell'INFN di Pisa, Pisa, Italy}\\
 F.~Costantini\orcid{0000-0002-2974-0067},
 N.~Doble$\,${\footnotemark[11]}\orcid{0000-0002-0174-5608},
 L.~Fiorini$\,${\footnotemark[18]}\orcid{0000-0002-5070-2735},
 S.~Giudici\orcid{0000-0003-3423-7981},
 G.~Pierazzini$\,$\renewcommand{\thefootnote}{\fnsymbol{footnote}}\footnotemark[2]\renewcommand{\thefootnote}{\arabic{footnote}}, 
 M.~Sozzi\orcid{0000-0002-2923-1465},
 S.~Venditti
\vspace{0.5cm}

{\bf Scuola Normale Superiore e Sezione dell'INFN di Pisa, Pisa, Italy}\\
 G.~Collazuol$\,${\footnotemark[19]}\orcid{0000-0002-7876-6124},
 L.~Di Lella$\,${\footnotemark[11]}\orcid{0000-0003-3697-1098},
 G.~Lamanna$\,${\footnotemark[20]}\orcid{0000-0001-7452-8498},
 I.~Mannelli\orcid{0000-0003-0445-7422},
 A.~Michetti
\vspace{0.5cm}

{\bf Sezione dell'INFN di Pisa, Pisa, Italy}\\
 C.~Cerri,
 R.~Fantechi\orcid{0000-0002-6243-5726}
\vspace{0.5cm}

{\bf DSM/IRFU -- CEA Saclay, Gif-sur-Yvette, France}\\
 B.~Bloch-Devaux$\,${\footnotemark[21]}\orcid{0000-0002-2463-1232},
 C.~Cheshkov$\,${\footnotemark[22]},
 J.B.~Ch\`eze,
 M.~De Beer,
 J.~Derr\'e, 
 G.~Marel,
 E.~Mazzucato,
 B.~Peyaud,
 B.~Vallage
\vspace{0.5cm}

{\bf Fachbereich Physik, Universit\"at Siegen, Siegen, Germany}\\
 M.~Holder,
 M.~Ziolkowski\orcid{0000-0002-2891-8812}
\vspace{0.5cm}

{\bf Dipartimento di Fisica dell'Universit\`a e Sezione dell'INFN di Torino, Torino, Italy}\\
 S.~Bifani\orcid{0000-0001-7143-8200},
 M.~Clemencic$\,${\footnotemark[15]}\orcid{0000-0003-1710-6824},
 S.~Goy Lopez$\,${\footnotemark[23]}\orcid{0000-0001-6508-5090}
\vspace{0.5cm}

{\bf Sezione dell'INFN di Torino, Torino, Italy}\\
 C.~Biino\orcid{0000-0002-1397-7246},
 N.~Cartiglia\orcid{0000-0002-0548-9189},
 F.~Marchetto\orcid{0000-0002-5623-8494}
\vspace{0.5cm}

{\bf Österreichische Akademie der Wissenschaften, Institut f\"ur Hochenergiephysik, Wien, Austria}\\
 H.~Dibon,
 M.~Jeitler\orcid{0000-0002-5141-9560},
 M.~Markytan,
 I.~Mikulec\orcid{0000-0003-0385-2746},
 G.~Neuhofer,
 L.~Widhalm$\,$\renewcommand{\thefootnote}{\fnsymbol{footnote}}\footnotemark[2]\renewcommand{\thefootnote}{\arabic{footnote}}
\vspace{0.5cm}

{\bf Authors affiliated with an international laboratory covered by a cooperation agreement with CERN}\\
 S.~Balev$\,$\renewcommand{\thefootnote}{\fnsymbol{footnote}}\footnotemark[2]\renewcommand{\thefootnote}{\arabic{footnote}},
 P.L.~Frabetti,
 E.~Gersabeck$\,${\footnotemark[24]}\orcid{0000-0002-2860-6528},
 E.~Goudzovski$\,${\footnotemark[1]}\orcid{0000-0001-9398-4237},
 P.~Hristov$\,${\footnotemark[15]}\orcid{0000-0003-1477-8414}, 
 V.~Kekelidze\orcid{0000-0001-8122-5065},
 A.~Korotkova$\,$\renewcommand{\thefootnote}{\fnsymbol{footnote}}\footnotemark[1]\renewcommand{\thefootnote}{\arabic{footnote}},
 V.~Kozhuharov$\,${\footnotemark[25]}\orcid{0000-0002-0669-7799},
 L.~Litov$\,${\footnotemark[25]}\orcid{0000-0002-8511-6883},
 D.~Madigozhin$\,$\renewcommand{\thefootnote}{\fnsymbol{footnote}}\footnotemark[1]\renewcommand{\thefootnote}{\arabic{footnote}}\orcid{0000-0001-8524-3455},
 N.~Molokanova,
 I.~Polenkevich,
 Yu.~Potrebenikov\orcid{0000-0003-1437-4129},
 S.~Stoynev$\,${\footnotemark[26]}\orcid{0000-0003-4563-7702},
 A.~Zinchenko$\,$\renewcommand{\thefootnote}{\fnsymbol{footnote}}\footnotemark[2]\renewcommand{\thefootnote}{\arabic{footnote}}
\vspace{0.5cm}

\end{raggedright}

%
%

\setcounter{footnote}{0}
\newlength{\basefootnotesep}
\setlength{\basefootnotesep}{\footnotesep}

\renewcommand{\thefootnote}{\fnsymbol{footnote}}
\noindent
$^{\footnotemark[1]}${Corresponding authors: A.~Korotkova, D.~Madigozhin, \\
email: anna.korotkova@cern.ch, dmitri.madigozhin@cern.ch}\\
$^{\footnotemark[2]}${Deceased}\\
\renewcommand{\thefootnote}{\arabic{footnote}}
$^{1}${Present address: School of Physics and Astronomy, University of Birmingham, Birmingham, B15 2TT, UK} \\
$^{2}${Also at Universit\`a degli Studi del Piemonte Orientale, I-13100 Vercelli, Italy} \\
$^{3}${Present address: Istituto di Cosmogeofisica del CNR di Torino, I-10125 Torino, Italy} \\
$^{4}${Present address: Sezione dell'INFN di Perugia, I-06100 Perugia, Italy} \\
$^{5}${Present address: Physics Department, University of Lancaster, Lancaster, LA1 4YW, UK} \\
$^{6}${Present address: School of Physics and Astronomy, University of Glasgow, Glasgow, G12 8QQ, UK} \\
$^{7}${Present address: Physics Department, Imperial College London, London, SW7 2BW, UK} \\
$^{8}${Present address: Centre de Physique des Particules de Marseille, Universit\'e Aix Marseille, CNRS/IN2P3, F-13288, Marseille, France} \\
$^{9}${Present address: University of California, Merced, CA 95344, USA} \\
$^{10}${Present address: George Mason University, Fairfax, VA 22030, USA} \\
$^{11}${Present address: Institut f\"ur Physik, Universit\"at Mainz, D-55099 Mainz, Germany} \\
$^{12}${Also at Dipartimento di Scienze Fisiche, Informatiche e Matematiche, Universit\`a di Modena e Reggio Emilia, I-41125 Modena, Italy} \\
$^{13}${Also at Istituto di Fisica, Universit\`a di Urbino, I-61029 Urbino, Italy} \\
$^{14}${Present address: SLAC National Accelerator Laboratory, Stanford University, Menlo Park, CA 94025, USA} \\
$^{15}${Present address: CERN, CH-1211 Gen\`eve 23, Switzerland} \\
$^{16}${Present address: National Center for Nuclear Research, Swierk, P-05-400, Poland} \\
$^{17}${Present address: Universit\`a di Roma La Sapienza, Roma, Italy} \\
$^{18}${Present address: Instituto de F\'isica Corpuscular IFIC, Universitat de Val\`encia, E-46071 Val\`encia, Spain} \\
$^{19}${Present address: Dipartimento di Fisica dell'Universit\`a e Sezione dell'INFN di Padova, I-35131 Padova, Italy} \\
$^{20}${Present address: Dipartimento di Fisica dell'Universit\`a e Sezione dell'INFN di Pisa, I-56100 Pisa, Italy} \\
$^{21}${Present address: Dipartimento di Fisica dell'Universit\`a, I-10125 Torino, Italy} \\
$^{22}${Present address: Institut de Physique Nucl\'eaire de Lyon, IN2P3-CNRS, Universit\'e Lyon I, F-69622 Villeurbanne, France} \\
$^{23}${Present address: Centro de Investigaciones Energeticas Medioambientales y Tecnologicas, E-28040 Madrid, Spain} \\
$^{24}${Present address: School of Physics and Astronomy, The University of Manchester, Manchester, M13 9PL, UK} \\
$^{25}${Present address: Faculty of Physics, University of Sofia, 1164 Sofia, Bulgaria} \\
$^{26}${Present address: Fermi National Accelerator Laboratory, Batavia, IL 60510, USA} \\

\newpage
\section{Introduction}
Semileptonic four-body decays of kaons, $K^{\pm} \rightarrow \pi^{\pm, 0} \pi^{\mp,0} l^{\pm}{\nu}$ ($K^{+-}_{l4}, K^{00}_{l4}$) with $l= e,\mu$,   are of particular interest because of the small number of hadrons in the final 
state, which allows studying low energy QCD,  
while the electroweak amplitude responsible for the leptonic part is well-understood in the Standard Model.
The development over more than 30 years of chiral perturbation theory (ChPT) \cite{Weinberg:1978kz, Bijnens:1994ie} has allowed predictions of form 
factors and decay rates at a precision level competitive with the accuracy of the experimental results
in the electron modes \cite{Batley:2010zza, NA482:2012cho, Batley:2014xwa}. 
In particular, the most recent theoretical works, prompted by the precise experimental measurements, have focused on form factor evaluation at higher orders, including radiative and isospin breaking effects, and developed a dispersive approach to match Low Energy Constants of ChPT~\cite{Bernard:2015vqa,Colangelo:2015kha}.
The muon modes are still to be investigated experimentally as 
the $\pi^+ \pi^-$ mode observation relies on a few events \cite{Bisi:1967zz} and the $\pi^0 \pi^0$ mode has not been observed.
This study  reports the first observation and branching ratio  (BR) measurement of the $K^{\pm} \rightarrow {\pi}^{0}{\pi}^{0}{\mu}^{\pm}{\nu}$ 
decay mode by the NA48/2 experiment at the CERN SPS.

\section{Theoretical framework and available measurements}
\label{Theory}
\begin{figure}[ht]
\center{\includegraphics[width=0.5\linewidth]{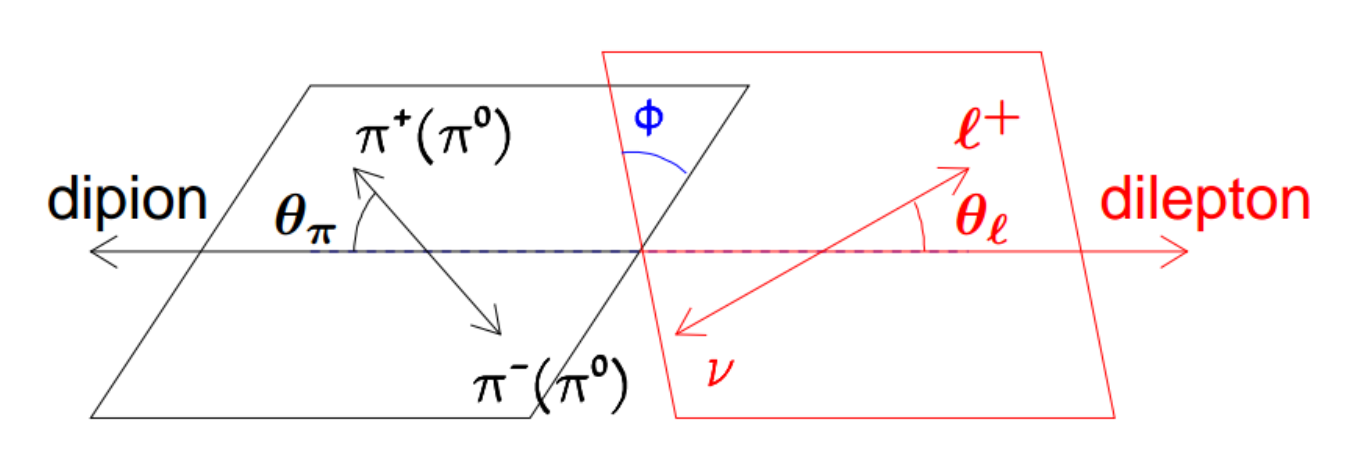}}
\caption{Cabibbo-Maksymowicz variables describing the semileptonic $K^{+-}_{l4} (K^{00}_{l4})$ decays.}
\label{fig:cama-image}
\end{figure}

 The differential rate of the $K^{00}_{l4}$ decay may be parameterized in terms of  
the five Cabibbo-Maksymowicz variables \cite{Cabibbo:1965zzb} illustrated in Fig.~\ref{fig:cama-image}:
$S_\pi$, the squared mass of the dipion system;
$S_l$, the squared mass of the dilepton system;
${\theta_{\pi}}$, the angle of a pion direction in the dipion rest frame with respect to the dipion line of flight in the kaon rest frame; 
$\theta_l$, 
the angle of the charged lepton direction in the dilepton rest frame
with respect to the dilepton line of flight in the kaon rest frame;
and $\phi$, the angle between the dipion and dilepton planes in the kaon rest frame. 

Unlike the $K^{+-}_{l4}$ mode, where the definition of $\cos\theta_{\pi}$ and $\phi$ is driven by the $\pi^{+}$ meson, the final state with two neutral pions cannot favour one particular $\pi^0$. Integration over the $\cos\theta_{\pi}$ and $\phi$ variables leads to the simplified differential rate~\cite{Bijnens:1994ie} 
\begin{align}
d\Gamma_3  &= \frac{G^2_F |V_{us}|^2 (1-z_{l})^2 \sigma_{\pi} X}{2^{11} \pi^5 m^5_{K^+}}  ( I_{1} + I_{2} ~\cos2\theta_ l +I_{6}~\cos\theta_l ) ~dS_\pi dS_l d\cos\theta_l,
\label{formula:dG3}
\end{align}
where 
$G_F$ is the Fermi coupling constant, 
$V_{us}$ is the Cabibbo-Kobayashi-Maskawa  matrix element,
$z_{l}=m^{2}_{l}/S_l$,
$\sigma_{\pi}=\sqrt{1-4m^{2}_{\pi^0}/S_\pi}$,
$X=\frac{1}{2}\sqrt{\lambda(m^{2}_{K^+}, S_\pi, S_l)}$,
and $\lambda(x,y,z) = x^{2}+y^{2}+z^{2}-2(xy+xz+yz)$ is the triangle function.  
The terms $I_{1}$, $I_{2}$, $I_{6}$ carry the dependence on  the kinematic variables ($S_\pi$, $S_l$) and combinations of the complex hadronic form factors $F_1$, $F_4$. In contrast to the electron mode, terms including $z_\mu$ cannot be neglected and contribute to the decay amplitude:
\begin{align}
I_1 = \{ (1+z_{\mu} )|F_1|^2 +2 z_{\mu}|F_4|^2 \}/4, 
~~I_2 = -(1- z_{\mu} )|F_1|^2 /4, 
~~I_6 = z_{\mu} \Re(F_{1}^{*}  F_4).
\end{align}
\noindent The two complex hadronic form factors $F_1, F_4$ are functions of the real form factors $F$ and $R$.
Considering the isospin decomposition of $F_1$, $F_4$, the $F$, $R$ form factors in the neutral pion $(00)$ mode are related to those of the charged pion mode $(+-)$ by:
$(F,R)_{00} = -(F^+ ,R^+ )_{+-}$,
where ($F^+ ,R^+) $ are the symmetric parts of $(F,R)$. 
Using the notations $P$ and $L$ of the four-vector sum of the two pions  and  the four-vector sum of the two leptons, respectively, the form factors are written as 
$F_1 = - X F$ and $F_4 = (PL) F + S_\mu R$ in the neutral pion mode, under the assumption of no isospin violating contributions $(m_u = m_d = \alpha_{QED} = 0)$.

The $F$ form factor has been measured precisely by NA48/2 in the $K_{e4}^{00}$ decay mode \cite{Batley:2014xwa} and may be used in the $K_{\mu4}^{00}$ decay assuming lepton flavour universality. The dependence of $F$ with $q^{2} = S_\pi/4m^{2}_{\pi^{+}}-1$ and  $S_l$ ($l= e,\mu$)  is 
\begin{equation}
 F(K^{00}_{l4}) = \left\{ \begin{array}{ll} 1+aq^2 + b q^4 + c \cdot S_l/4m_{\pi^{+}}^2 , & q^2 \ge 0 \\ 
 1 + d\sqrt{|q^2/(1+q^2)|} + c\cdot S_l/4m_{\pi^{+}}^2, & q^2 \le 0 \end{array} \right.
 \label{form:Fke4}
 \end{equation}
  where 
  $a =   0.149 \pm 0.033 \pm 0.014$,
  $b = - 0.070 \pm 0.039 \pm 0.013$,
  $c =   0.113 \pm 0.022 \pm 0.007$,
  $d = - 0.256 \pm 0.049 \pm 0.016$. 
  In each case, the first error quoted is statistical and the second error is systematic.
  The $F$ absolute normalization has been also measured as $F = f \cdot F(K^{00}_{e4})$,
where $f = 6.079 \pm 0.012_{\rm stat} \pm 0.027_{\rm syst} \pm 0.046_{\rm ext}$.

In contrast, the $R$ form factor, which does not contribute to $K_{e4}$ decays, has never been measured so far and only theoretical calculations exist at various orders of ChPT \cite{Bijnens:1994ie}.
\section{Beams and detectors}
\label{Beams_detectors}

\begin{figure}[t]
\center{\includegraphics[width=1.
\linewidth]
{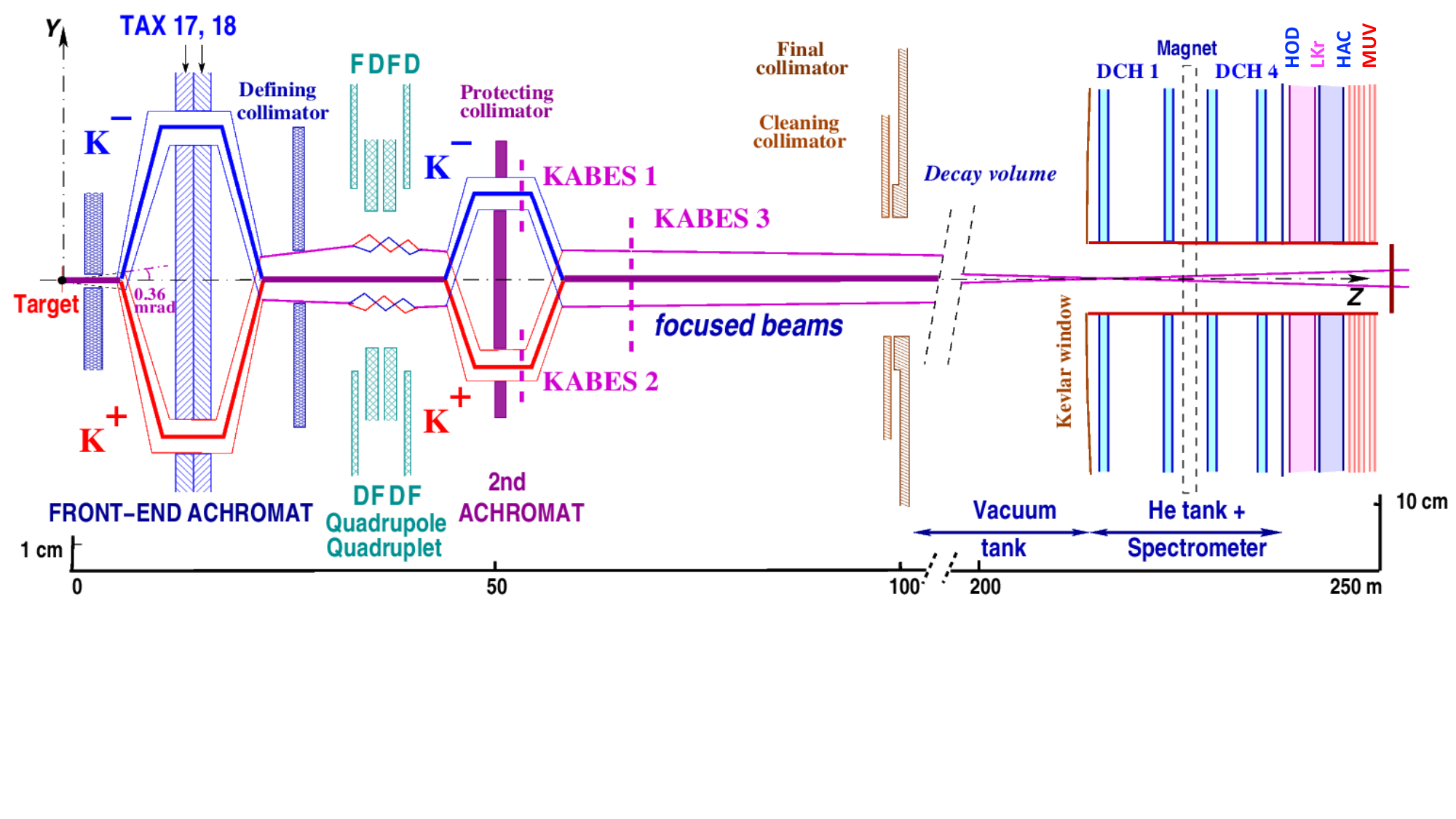}
}\vspace{-25mm}
\caption{Schematic side view of the NA48/2 setup.}
\label{ris:kabes_sh1}
\end{figure}

The NA48/2 experiment at the CERN SPS was designed to search for direct CP violation in $K^\pm$ decays to three pions~\cite{NA482:2007ucr}. 
During the 2003--2004 data taking period, the 400 GeV proton beam from the SPS was impinging on a beryllium target to produce simultaneous $K^+$ and $K^-$ beams 
(Fig.~\ref{ris:kabes_sh1}).
A front-end achromat separated locally positively and negatively charged beams in the vertical direction, allowing a dump-collimator (TAX 17, 18) with two holes, 20~cm apart, to select positive and negative particles with a central momentum of 60~GeV/$c$ and a momentum band of 3.8\% (rms).
 A second achromat separated again locally the positive and negative beams by 8~cm, allowing the three stations of the kaon beam spectrometer (KABES)~\cite{Peyaud:2004tj} to measure the momentum of each individual particle. 
Each KABES station consisted of two time projection chambers using the MICROMEGAS technology \cite{Giomataris:1995fq}.
 The achieved resolutions were 800~$\mu$m, better than 1\% and 600~ps for space point, momentum and time, respectively.
At the exit of this achromat the beams traveled on a common axis towards the decay region. The two resulting beams, each 1~cm wide in the transverse plane, were superimposed in the decay volume  enclosed in a 114 m long vacuum tank.

Charged products from the $K^{\pm}$ decays were measured by a magnetic spectrometer consisting of four drift chambers (DCH) and
a dipole magnet located between the second and the third chamber. The spectrometer was located in a tank filled with helium
at atmospheric pressure and separated from the decay volume by a thin Kevlar composite window.
The magnet provided a transverse momentum kick of $\Delta p$ = 120 MeV/$c$ to charged particles in the horizontal plane. The spatial resolution of each DCH was $\sigma_x$ = $\sigma_y$ = 90~$\mu$m and the achieved momentum resolution was $\sigma_p / p = (1.02 \oplus 0.044 \cdot p)$\% ($p$ in GeV/$c$).
The spectrometer was followed by a hodoscope (HOD) consisting 
of two planes of plastic scintillators segmented into horizontal and vertical strips and arranged in four quadrants.

A liquid krypton calorimeter (LKr) located behind the HOD was used to reconstruct $\pi^{0} \rightarrow \gamma \gamma$ decays.
It was an almost homogeneous ionization chamber with an active volume of about $10$~m$^3$ of liquid krypton, segmented
transversely into 13248 projective cells, $2 \times 2$~cm$^2$ each.
The calorimeter energy resolution was $\sigma_E /E = (3.2/\sqrt{E} \oplus 9/E \oplus 0.42)$\%  ($E$ in GeV).

The LKr was followed by a hadronic calorimeter (HAC) with a total iron thickness of 1.2~m. 
A Muon Veto system (MUV), consisting of three layers made of 80 cm iron followed by scintillators, was used to identify muons.
A detailed description of the NA48 detector and its performance has been published in \cite{Fanti:2007vi}.

The experiment collected a total of $1.2 \times 10^{10}$ triggers in two years of data-taking using a dedicated two-level trigger logic to select and flag events. In the  study reported here, a sub-sample of $2 \times 10^8$ triggers is considered: the first level trigger required a signal in at least one HOD quadrant 
in coincidence with the presence of energy deposits in LKr consistent with at least two photons.
At the second level,  
an on-line processor receiving the DCH information reconstructed the momentum of charged particles and calculated the missing mass  
to the $(K^{\pm} - \pi^{\pm})$ system under the assumption that the particles were $\pi^{\pm}$ originating from the decay of a 60 GeV/$c$ $K^{\pm}$ traveling along the nominal beam axis. 
The requirement that  the missing mass 
exceeds the $\pi^0$ mass was imposed to reject most $K^{\pm} \to \pi^{\pm} \pi^0$ decays (the lower trigger cutoff was 194 MeV/$c^2$ in 2003 and 181 MeV/$c^2$ in 2004).

\section{Event reconstruction and selection}
\label{Selection}

The signal branching ratio BR$(K^{00}_{\mu 4})$ is measured relative to the abundant normalization channel $K^{\pm} \rightarrow {\pi}^{0}{\pi}^{0}{\pi}^{\pm}$ ($K^{00}_{3\pi}$),  
which leads to identical numbers of detected charged and neutral particles and is collected concurrently through the same trigger logic: 
 \begin{equation}
   {\rm BR}(K^{00}_{\mu 4}) = \frac{N_{S}}{N_{N}} \cdot \frac{A_{N}}{A_{S}} \cdot K_{\rm trig} \cdot {\rm BR}(K^{00}_{3\pi}),
   \label{BranchingFull}
\end{equation}
where $N_{N}$ and $N_{S}$ are the numbers of selected events after background subtraction in the normalization and signal samples, respectively;  the corresponding selection
acceptances $A_{N}$ and $A_{S}$ are computed using a detailed GEANT3-based~\cite{Brun:1994aa} Monte Carlo simulation (MC); and $K_{\rm trig}$ is a  
factor accounting for possible differences in the  trigger efficiency between the signal and normalisation samples.
The kinematic cut applied at the second trigger level is fully efficient for both $K^{00}_{\mu 4}$ and $K^{00}_{3\pi}$ decays. The normalization branching ratio is BR$(K^{00}_{3\pi}) = {\ensuremath{(1.760 \pm 0.023)}\%\xspace}$~\cite{ParticleDataGroup:2022pth}. 

A common selection is considered,  followed by exclusive criteria to distinguish normalization and signal candidates.
The common selection includes  a charged and a neutral selection.

The charged selection requires
a DCH track with momentum $P_{\rm DCH}$ in the range \mbox{5--35~GeV/$c$} and  the distance between the track and the beam axis in the DCH1 plane larger than 12~cm;
at least one KABES track with momentum $P_{\rm KABES}$ in the range \mbox{54--67~GeV/$c$}, within 10~ns  of the DCH track time and  carrying the same charge as the DCH track.
The best matching KABES track and the DCH track are propagated to the closest point of approach to define the charged vertex with longitudinal position $Z_{\rm c}$.

The neutral selection requires
four photon candidates defined as energy deposits in the LKr calorimeter larger than 3~GeV/$c$,
separated by at least 2~cm from any inactive cell,
by at least 10~cm from any other photon candidate in-time within 5~ns, and by at least 
15~cm from the  extrapolated position at the LKr front plane of any DCH track with an associated HOD time within 10~ns of any photon candidate.
Each of the photon candidates should be within 2.5~ns of the four-photon average time.  The four candidates should be consistent with the decay of two neutral pions
at longitudinal positions $Z_1$ and $Z_2$, 
obtained under the assumption that each pair of photons is produced by a $\pi^{0} \rightarrow \gamma\gamma$ decay,
with $| Z_{1} - Z_{2} | < 500$~cm.
The neutral vertex position is defined as the average of the two  $\pi^{0}$ vertex positions:
$Z_{\rm n}= (Z_{1}+Z_{2})/2$, 
and  is required to be in a 106~m long decay region upstream of DCH1.

Combinations with $|Z_{\rm n} - Z_{\rm c}| < 600$~cm are considered further. In case several combinations satisfy this condition, the one with the smallest discriminant value is selected. The discriminant $D_{\rm nc}$ takes into account the resolution of  $Z_{1}-Z_{2}$ and $Z_{\rm n}-Z_{\rm c}$ according to
\begin{equation}
\begin{split}
D_{\rm nc} = \left(\frac{Z_{1}-Z_{2}}{\sigma_{12}(Z_{\rm n})}\right)^{2}+\left(\frac{Z_{\rm n}-Z_{\rm c}}{\sigma_{\rm nc}(Z_{\rm av})}\right)^{2},
\end{split} 
\end{equation}
where $\sigma_{12}, \sigma_{\rm nc}$ are the Gaussian widths of the $Z_{1}-Z_{2}$ and $Z_{\rm n}-Z_{\rm c}$ distributions, measured on data and parameterized as a function of $Z_{\rm n}$ 
and $Z_{\rm av} = (Z_{\rm n}+Z_{\rm c})/2$, respectively.

\begin{figure}[ht] 
\includegraphics[width=1.\linewidth]{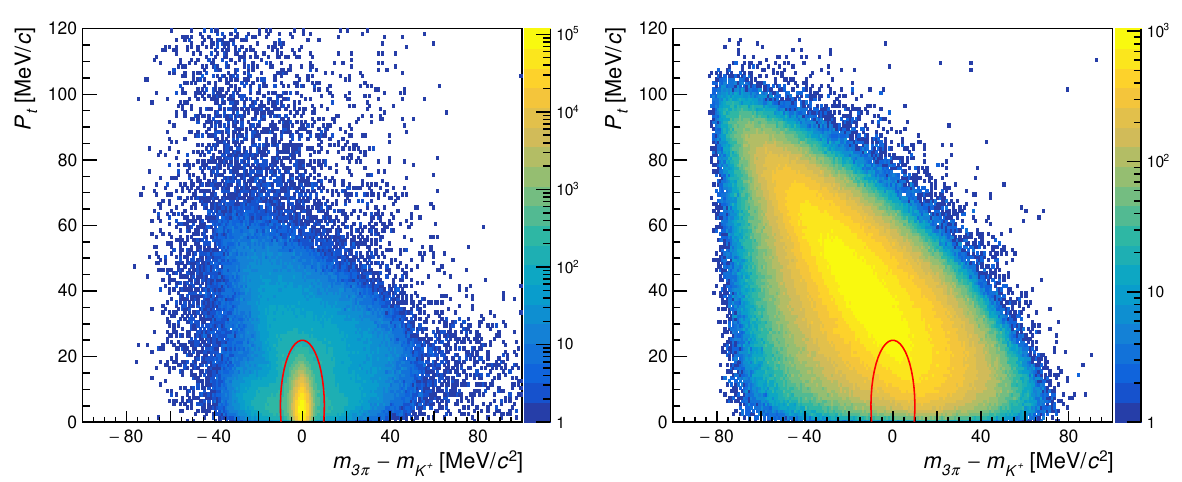} 
\caption{Distribution of simulated events in the reconstructed $(m_{3\pi}-m_{K^+}, P_t)$ plane after 
the common selection, for $K^{00}_{3\pi}$ decays (left) and  $K^{00}_{\mu4}$ decays (right). Events within the ellipse are selected as normalization candidates. Events outside the ellipse are selected as signal candidates.
}
\label{fig:MCEllipses}
\end{figure}

Each candidate is reconstructed in the $(m_{3\pi}, P_t)$ plane, where $m_{3\pi}$ is the mass of the three pion system 
(assuming a $\pi^+$ mass for the DCH track) and $P_t$ is its transverse momentum relative to the nominal beam axis.
The $K^{00}_{3\pi}$ normalization candidates  are required  
to be inside an ellipse centred at the kaon mass and $P_t$ value of 5 MeV/$c$, with semi-axes 10 MeV/$c^2$ and 20 MeV/$c$, respectively, thus ensuring fully reconstructed $K^{00}_{3\pi}$ three-body decays (Fig.~\ref{fig:MCEllipses}).
      
The selection of $K^{00}_{\mu4}$ signal candidates requires the reconstructed $m_{3\pi}$ and $P_t$ values 
to be outside the $K^{00}_{3\pi}$ ellipse. Additional conditions are applied to identify a muon in the final state and suppress the contribution of  $K^{00}_{3\pi}$  with a decay in flight  $\pi^{\pm} \to \mu^{\pm} \nu$ which mimics the signal kinematics.
Muon identification requires the deflected DCH track, extrapolated  to the MUV plane, to be within the instrumented area 
and  spatially associated to a MUV signal. A condition $P_{\rm DCH} > $ 10~GeV/$c$ is applied to ensure high muon identification efficiency.

The missing mass squared variable $m^2_{\rm miss}$ used to discriminate signal and background candidates is evaluated using the 4-momenta of all measured particles:
\begin{equation}
m^2_{\rm miss}=(E_{K}-E_{\pi_{1}^{0}}-E_{\pi_{2}^{0}}-E_{\mu})^{2}-(\vec{P}_{\rm KABES}-\vec{P}_{\pi_{1}^{0}}-\vec{P}_{\pi_{2}^{0}}-\vec{P}_{\rm DCH})^{2},
\label{form:Mmiss}
\end{equation}
where $E_{K} = \sqrt{P_{\rm KABES}^2 + m^2_{K^+}}$  and $E_{\mu} = \sqrt{P_{\rm DCH}^2 + m^2_{\mu}}$ are the kaon and muon energies, respectively.
In a similar way, $m^2_{\rm miss}(\pi)$ is defined as a squared 
missing mass reconstructed assigning the charged pion mass $m_{\pi^+}$ to the DCH track instead of $m_{\mu}$.

\begin{figure}[ht]
\includegraphics[width=1.\linewidth]{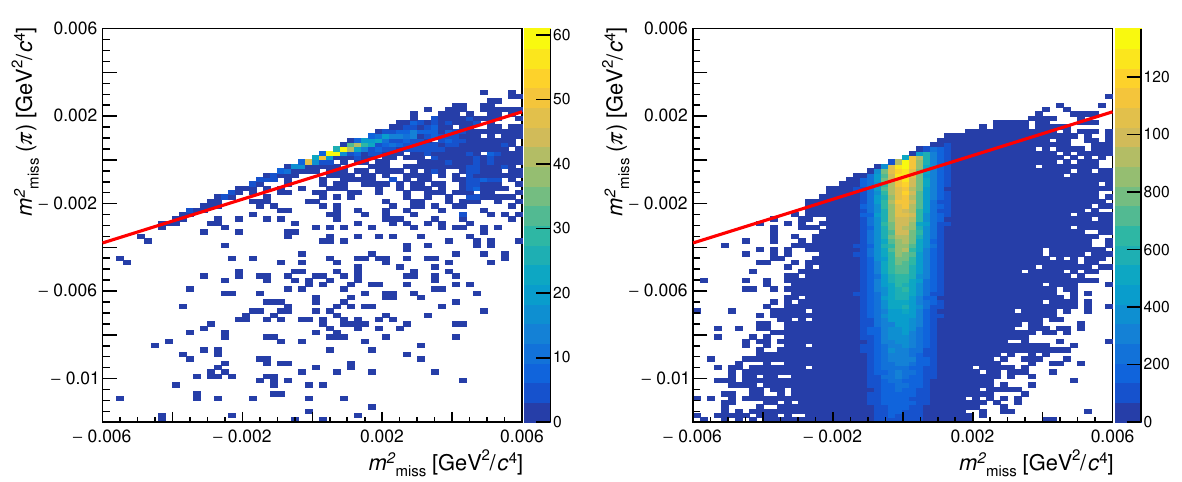} 
\caption{Distributions of simulated $K^{00}_{3\pi}$ background (left) and $K^{00}_{\mu4}$ signal candidates (right) in the reconstructed ($m^2_{\rm miss}$, $m^2_{\rm miss}(\pi)$) plane after the signal selection, 
but before applying the condition in this plane. Candidates above the line are excluded.
}
\label{fig:PiMM2cut}
\end{figure}

Values of $m^2_{\rm miss}(\pi)$ close to zero correspond to a $K^{00}_{3\pi}$ decay kinematics. 
Due to the correlation between the reconstructed $m^2_{\rm miss}(\pi)$ and $m^2_{\rm miss}$ values,
a selection condition is applied in the $(m^2_{\rm miss}(\pi)$, $m^2_{\rm miss})$ plane, $m^2_{\rm miss}(\pi) < {0.5\xspace} m^2_{\rm miss} - {0.0008\xspace}~\textrm{GeV}^2/c^4$, to reject the $K^{00}_{3\pi}$  decays in the signal sample (Fig. \ref{fig:PiMM2cut}).
\begin{figure}[h]
  \centering
\includegraphics[width=1.\linewidth]{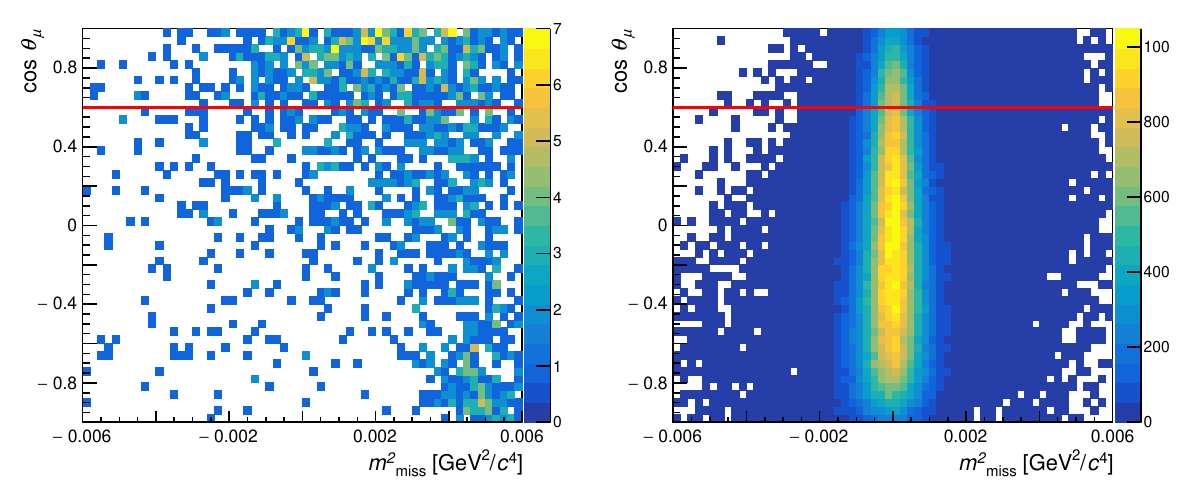}
  \caption{Distribution of simulated  $K^{00}_{3\pi}$ background (left) and $K^{00}_{\mu4}$ signal candidates (right) in the reconstructed $(m^2_{\rm miss}, \cos\theta_{\mu})$ plane after 
  the signal selection, 
  but before applying the condition in this plane. 
  Candidates above the line are excluded.
 }
   \label{fig:cosThetaMu}
\end{figure}
\begin{figure}[ht]
   \centering
\includegraphics[width=1.\linewidth]{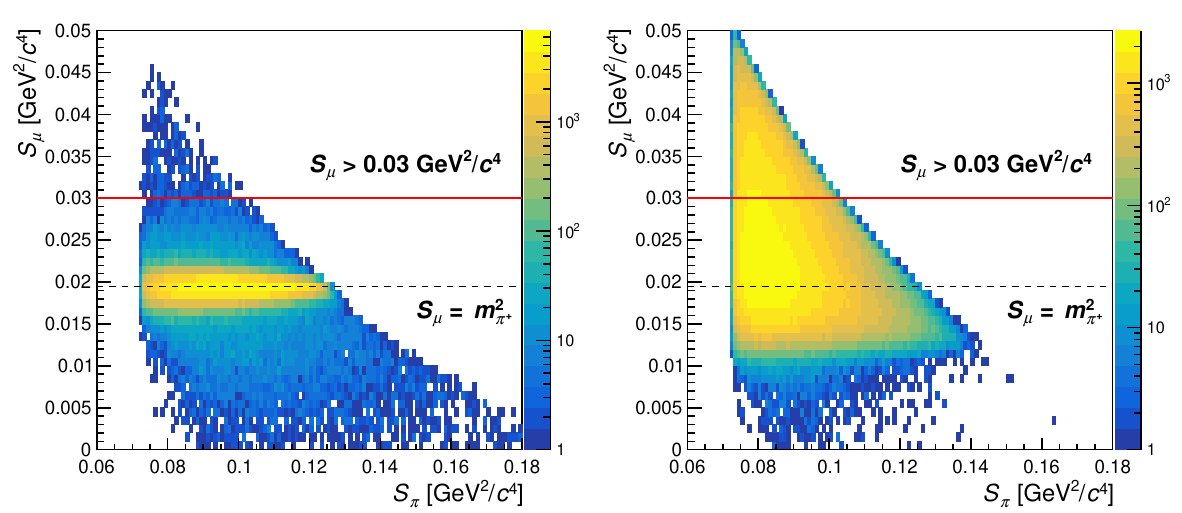} 
\caption{Distributions of simulated $K^{00}_{3\pi}$ background (left) and $K^{00}_{\mu4}$ signal candidates (right) in the reconstructed ($S_\pi$, $S_\mu$) plane after the signal selection, 
  but before applying the condition in this plane. Candidates above the solid line are selected.
  }
   \label{fig:ShowCutSl}
\end{figure}
\clearpage

Residual background  from the $K^{00}_{3\pi}$ decay followed by $\pi^\pm \to \mu^\pm \nu$ decay is suppressed by requiring $\cos\theta_{\mu} < 0.6$ 
(Fig.~\ref{fig:cama-image} with $l=\mu$), as expected from the different shapes of the signal 
and  $K^{00}_{3\pi}$ background distributions 
(Fig.~\ref{fig:cosThetaMu}).
The background is suppressed further by requiring that the reconstructed dilepton mass, computed as the squared missing mass to the dipion system, satisfies  the condition $S_\mu >{0.03\xspace}$~GeV$^2$/$c^4$ 
(Fig. \ref{fig:ShowCutSl}). The $K^{00}_{3\pi}$ events reconstructed with $S_\mu >{0.03\xspace}$~GeV$^2$/$c^4$ are due to possible mis-reconstruction of the track when the pion decays in the spectrometer, or an association to a mis-reconstructed  $\pi^0$ pair, as shown by simulations.

Applying the signal selection to the data sample leads to {3718\xspace} $K^{00}_{\mu 4}$ signal candidates, of which 2437 lie in the signal region $|m^2_{\rm miss}| < 0.002$~GeV$^2/c^4$. 
The regions $|m^2_{\rm miss}| > 0.002$~GeV$^2/c^4$ are used as control regions in the background evaluation.
The normalization channel conditions select ${\ensuremath{7.3 \times 10^{7}}\xspace}$ 
$K^{00}_{3\pi}$ reconstructed data events.
The normalization sample is considered as background-free. A possible relative contamination from $K^{00}_{e4}$ and $K^{00}_{\mu4}$ decays is below $10^{-4}$ and gives 
a negligible contribution to the uncertainty on the signal BR measurement. 
 \begin{figure}[t]
    \centering
\includegraphics[width=0.76\linewidth]{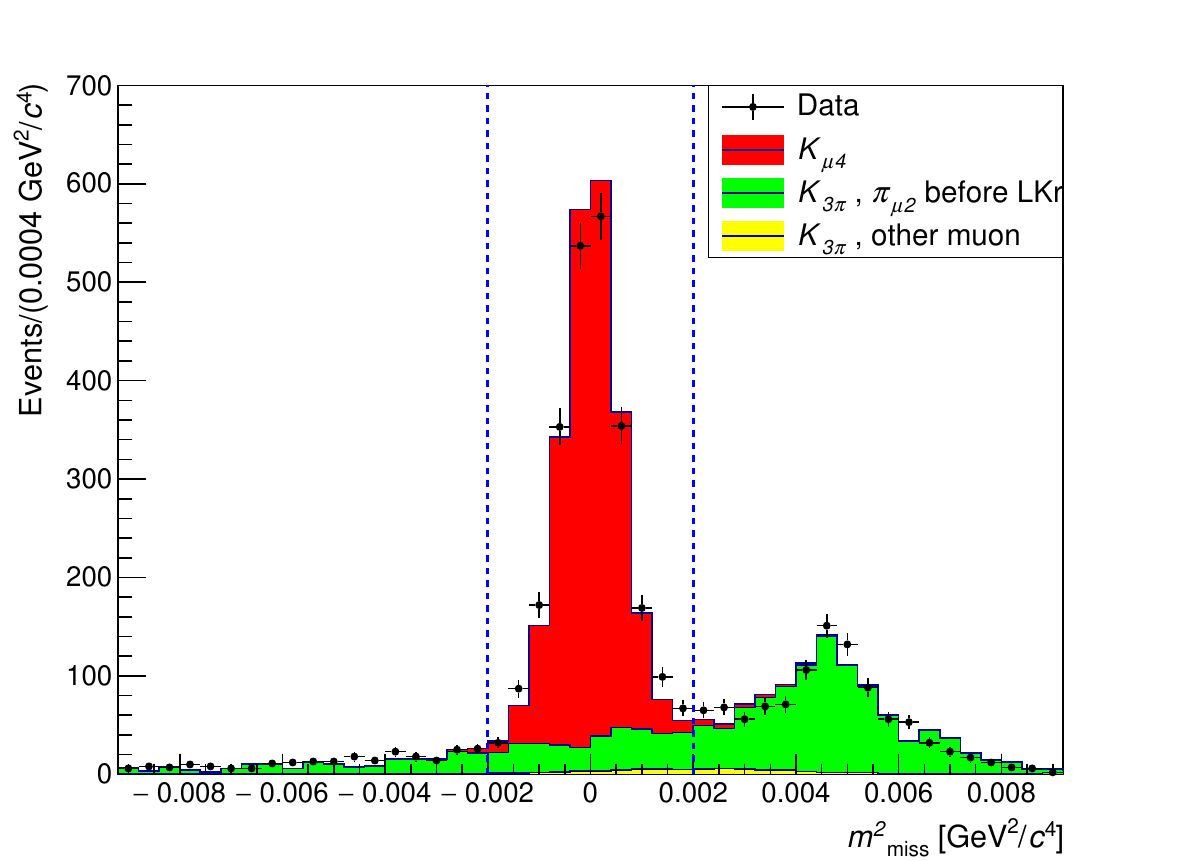} 
\caption{Distribution of $m^2_{\rm miss}$ for the selected data events (markers), simulated background  and signal contributions (histograms). 
    Vertical lines show the signal region.
    } 
\label{AllBackgrounds}
  \end{figure}

\section{Background evaluation}
\label{Background}

The background from $K^{00}_{3\pi}$ decays followed by  pion decays   upstream of the LKr front plane  is evaluated by simulation. The background from $K^{00}_{3\pi}$ decays with the  pion decaying or interacting in the LKr calorimeter and leading to a muon signal is evaluated from control data samples.
A method was developed 
to predict the $m^2_{\rm miss}$  shape of these contributions, 
using the pion track measured in the DCH spectrometer.
 
A control data sample is obtained  
by applying the $K^{00}_{\mu 4}$ signal selection, excluding the requirement of an associated MUV signal  and requiring a pion-like energy deposit  $E_{\rm LKr}$  in the LKr calorimeter such that $0.3 < E_{\rm LKr}/P_{\rm DCH} <0.8$.
This sample corresponds to secondary pions produced in  hadronic interactions and decaying to low-energy muons.
Another sample is obtained by weighting each event by the momentum-dependent decay probability, which corresponds to energetic pions decaying downstream of the LKr front plane before interacting.

 The $m^2_{\rm miss}$ spectra of data events, and those of simulated signal and estimated backgrounds, are shown in Fig.~\ref{AllBackgrounds}.
 The $K^{00}_{\mu 4}$ signal is observed as a peak in the $m^2_{\rm miss}$  distribution.
The background contributions are obtained from a fit of the two distributions to the data in the control $m^2_{\rm miss}$ region, excluding the signal region and
taking into account the simulated signal tails.
The contribution from $K^{00}_{3\pi}$ decays with the  pion decaying or interacting in the LKr calorimeter, labeled  
``other muon"  in Fig.~\ref{AllBackgrounds},
is found to be ten times smaller than the main background component. The number of background events in the signal region is found to be  
${\ensuremath{354 \pm 33}\xspace}_{\rm stat}$
by integrating the background contributions.

\section{Acceptances}
\label{MainGenerator}
The MC simulation is used to compute the selection acceptance for signal and normalization channels. The simulation includes full detector geometry and material description, detector local inefficiencies and misalignment, accurate simulation of the kaon beam line and time variations of the above throughout the data taking period.
 The muon identification efficiency is emulated according to the measurement using a $K^{00}_{3\pi}$ data sample selected by 
inverting the neutral and charged vertex matching condition and requiring $S_{\mu}$ to be consistent with $m_{\pi^+}^2$, thus  
ensuring a \mbox{$\pi^{\pm} \to \mu^{\pm} \nu$} decay.
The measured efficiency is parameterized as a function of the track momentum and distance from the beam axis in the MUV plane. Applying this model to simulated signal events, the integrated MUV inefficiency is found to be 1.65\%.

The signal channel $K_{\mu 4}^{00}$ is simulated  according to  
\cite{Bijnens:1994ie} including the {\em 1-loop} description of the form factor $R$ and the form factor $F$ measured in the $K_{e4}^{00}$  mode (Eq.~\ref{form:Fke4}). 
The signal region definition  $|m^2_{\rm miss}| < 0.002$ GeV$^2/c^4$
contains {98.2\%\xspace} of the selected MC events 
(Fig.~\ref{AllBackgrounds}).
The resulting acceptance is $A_S =   {\ensuremath{(3.453 \pm 0.007_{\rm stat})}\%\xspace}$ in the restricted kinematic  space $S_\mu >{0.03\xspace}$~GeV$^2$/$c^4$
and is $A_S = {\ensuremath{(0.651 \pm 0.001_{\rm stat})}\%\xspace}$ in the full  kinematic space.

The normalization channel $K_{3\pi}^{00}$  is generated  using the measured decay amplitude \cite{Batley:2000zz} implemented using an empirical parameterization of the data \cite{NA482:2010gwp}.
 The normalization acceptance is evaluated as  $A_N = {\ensuremath{(4.477 \pm 0.002_{\rm stat})}\%\xspace}$. The statistical uncertainties quoted are related to the sizes of the simulated samples and have a negligible impact on 
 the measurement.

\section{Systematic uncertainties and results}
\label{Results}
 The last ingredient needed by Eq.~(\ref{BranchingFull}) is the trigger correction $K_{\rm trig}$.
 Because of the similar topology of the signal and normalization decay modes, most inefficiencies, measured using control triggers from data or MC emulations, cancel at first order and lead to $K_{\rm trig}  = {\ensuremath{0.999 \pm 0.002}\xspace}$. 
The uncertainty in the measured $K_{\rm trig}$ factor is propagated as a systematic contribution.

The systematic uncertainty related to the background evaluation is estimated by considering alternative  
methods:
neglecting the smaller of the two background  components; 
restricting the control region to $m^2_{\rm miss}>0.004$~GeV$^2/c^4$; and adding extra smearing to the signal distribution. The largest deviation from the reference value is quoted as a systematic uncertainty, leading to $354 \pm 33_{\rm stat} \pm 62_{\rm syst}$ background events in the signal region.

The systematic uncertainties related to accidental in-time signals
are estimated by enlarging by a factor of two the time windows used for coincidence of KABES and DCH tracks, consistency of HOD track time and LKr photon times, and of the four LKr photon candidates.
The contribution is conservatively calculated as a linear sum of the unsigned shifts induced by the extension of each considered time window.

The systematic uncertainty related to the MUV inefficiency modelling in the simulation is 
quoted as 
20\% of its effect on the signal acceptance calculation,
which reflects its maximum variation when applying  tighter and looser selection conditions of the $K_{3\pi}^{00}$ data sample used in the evaluation.

The signal acceptance depends on the form factor description used in the simulation (Section~\ref{Theory}). The related uncertainty is estimated by weighting the simulated events according to the corresponding variations of the differential rate. Several modifications are investigated including replacing the description of $R_{\mathit 1-loop}$ by $R_{\mathit tree}$  \cite{Bijnens:1994ie};  modifying the relative contribution of $F$ and $R$ by 20\% at the reference point ($S_\pi = 4 m_{\pi^+}^2, S_\mu= m_{\mu}^2$); varying each parameter of $F(K^{00}_{e4})$ within its uncertainty. Conservatively, all observed variation are added in quadrature. 

Stability checks, considering data sub-samples defined by the kaon beam charge and data taking period  do not reveal any evidence for residual systematic effects.

Table~\ref{tab:Results} summarizes the achieved measurement in the kinematic space $S_\mu > {0.03\xspace}$~GeV$^2/c^4$,  BR$(K^{00}_{\mu 4}) = ({0.65\xspace} \pm {0.02\xspace}_{\rm stat} \pm {0.02\xspace}_{\rm syst} \pm {0.01\xspace}_{\rm ext}) \times 10^{-6}$,
together with the detailed uncertainties.
The  branching ratio measurement precision is dominated by the statistics of the signal sample and the uncertainty in the background evaluation.
As expected, the branching ratio in the full kinematic space, 
{\rm BR}$(K^{00}_{\mu 4})= (3.45 \pm {0.10\xspace}_{\rm stat} \pm {0.11\xspace}_{\rm syst} \pm {0.05\xspace}_{\rm ext}) \times 10^{-6}$, 
is more sensitive to the form factor modelling than the branching ratio in the restricted space: the corresponding uncertainty is 1.37\% , leading to a total systematic uncertainty of 
3.30\% and a total error of 4.55\%. All other components are scaled according to their relative  contributions $\delta{\rm  BR/BR}$.

\begin{table}[t]
\caption{Result of the {\rm BR}$(K^{00}_{\mu 4})$ measurement in the restricted kinematic space~\mbox{$S_\mu > {0.03\xspace}$~GeV$^2 /c^4$} and contributions of the considered uncertainties.}\label{tab:Results}
\centering
\vspace{0.7mm}
\begin{tabular}{| l | c | c |}  \hline
  {\rm BR}$(K^{00}_{\mu 4})$ central value $[10^{-6}]$  &   \multicolumn{2}{c|}{{0.651\xspace}}                       \\ \hline
                                        &       $\delta {\rm BR} [10^{-6}]$      &  $\delta {\rm BR/BR}$          \\ \hline
  Data statistical error                      &   {0.019\xspace}           &  {2.85\%\xspace}       \\  \hline
  MC   statistical error                      &   {0.001\xspace}         &  {0.21\%\xspace}     \\
  Trigger                               &   {0.001\xspace}       &  {0.18\%\xspace}   \\
  Background                            &  {0.019\xspace}        &  {2.96\%\xspace}    \\
  Accidentals                           &   {0.002\xspace}          &  {0.32\%\xspace}      \\
MUV inefficiency                      &  {0.002\xspace}          &  {0.33\%\xspace}      \\
  Form factor modelling                &  {0.001\xspace}     &  {0.14\%\xspace} \\ \hline 
  Total systematic error &   {0.020\xspace}        &  {3.01\%\xspace}    \\
  \hline
  BR$(K^{00}_{3\pi})$ error (external)        &    {0.009\xspace}         &  {1.31\%\xspace}     \\ \hline          
Total error                           &    {0.028\xspace}            &  {4.35\%\xspace}  \\ \hline
\end{tabular}

\end{table}

\begin{figure}[t]
        \centering 
\includegraphics[width=1.0\linewidth]{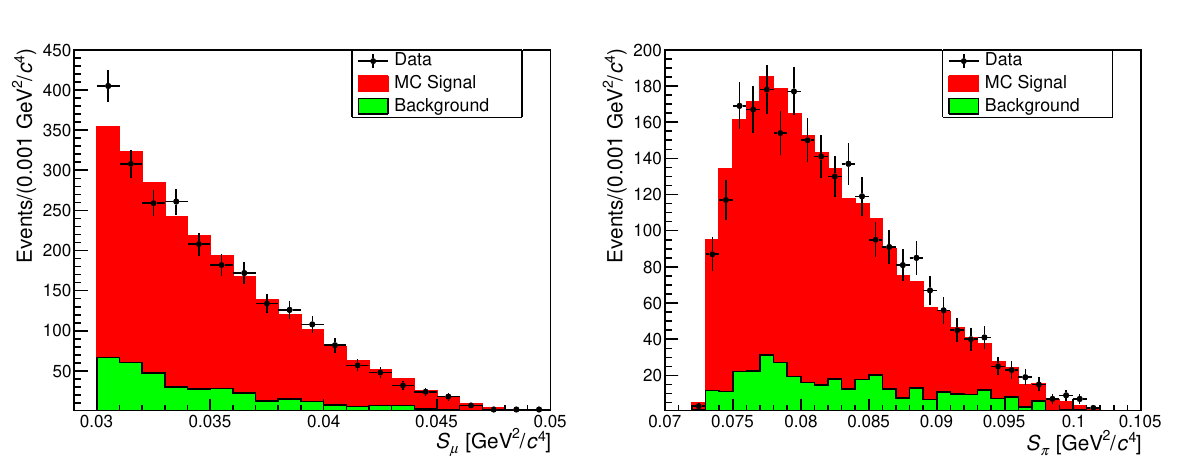} 
\caption{Distributions of $S_\mu$ and $S_\pi$ variables for data (markers), signal simulation and background estimation (histograms) for $S_\mu  > {0.03\xspace}$~GeV$^2 /c^4$.}
\label{fig:OneDimComp}
\end{figure}

The $S_\pi$ and $S_\mu$ distributions of the selected events  are shown in Fig.~\ref{fig:OneDimComp}  together with 
simulated signal and backgrounds.
The limited kinematic space accessible does not allow a  measurement of the $R$ form factor, while the observed agreement between data and simulation confirms a reasonable quality of the model used for the signal acceptance calculation.

\begin{figure}[!h]
        \centering     
\includegraphics[width=0.9\linewidth]{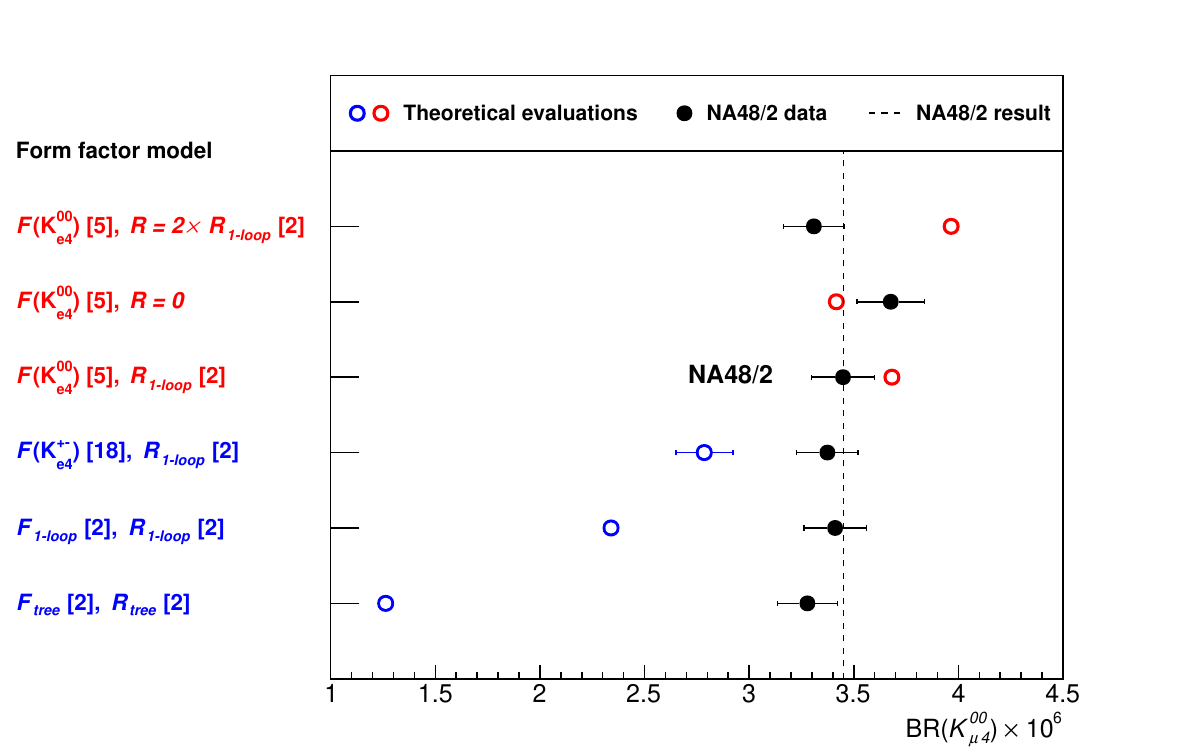} 
        \caption{
Evolution of the theoretical  evaluation of the branching fraction  BR($K^{00}_{\mu4}$) with the form factor model considered (open markers) and comparison with the NA48/2 measurement using the corresponding model in the selection acceptance  calculation (solid markers). The theoretical evaluations obtained using $F$ and $R$ form factors from ChPT  in~\cite{Bijnens:1994ie} are labeled as blue open markers.
The theoretical evaluations obtained by replacing the $F$ description from~\cite{Rosselet:1976pu} 
 by the more precise measurement of~\cite{Batley:2014xwa} 
are labeled as red open markers.
The NA48/2 measurement is most consistent with the theoretical evaluation considering the ChPT formulation $R_{1-loop}$.
}     \label{fig:FinalCompareNew}
      \end{figure}

As the signal acceptance depends on the form factor model considered,  the same data are used to  extract the branching ratio under  different assumptions.  The comparison of the  NA48/2 measurement with the corresponding theoretical predictions is shown in Fig. \ref{fig:FinalCompareNew}.  
The three lower predicted values of BR($K^{00}_{\mu4}$) 
correspond to the \textit{tree} level, \textit{1-loop} and ``beyond \textit{1-loop}'' models of the form factors $F$ and $R$, respectively~\cite{Bijnens:1994ie}. The ``beyond \textit{1-loop}'' model uses $R_{\mathit 1-loop}$ and includes the $F$ form factor measurement \cite{Rosselet:1976pu} obtained by the S118 experiment at the CERN PS 
using a sample of 30 000 $K_{e4}^{+-}$ decays. 
The other three values 
correspond to our evaluation of the predicted branching ratio using the $F(K^{00}_{e4})$ measurement of NA48/2 \cite{Batley:2014xwa} and three models of $R$: $R = R_{\mathit 1-loop}$ from~\cite{Bijnens:1994ie}, $R=0$ and $R = 2 \times R_{\mathit 1-loop}$, respectively.    The agreement of the measured branching ratio and the predicted value improves when introducing more elaborate models of the form factors. 
The data do not support the simple models of $F$ and $R$ form factors (\textit{tree} and \textit{1-loop} models) and are in agreement with the most recent $F$ measurement and $R_{\mathit 1-loop}$ calculation. 
The achieved NA48/2 measurement is 
not precise enough to determine the variation of $R$ with $S_\pi$ and $S_\mu$.
\section{Conclusion}
The NA48/2 experiment at CERN reports the first observation of the $K^{\pm} \to \pi^0 \pi^0 \mu^{\pm} \nu$ decay from a sample of {2437\xspace} signal candidates with 15\% background contamination.
Measurements of the branching ratio are obtained in the kinematic region $S_\mu > 0.03$~GeV$^2/c^4$ as
\mbox{BR$(K^{00}_{\mu 4}) = ({0.65\xspace} \pm {0.02\xspace}_{\rm stat} \pm {0.02\xspace}_{\rm syst} \pm {0.01\xspace}_{\rm ext}) \times 10^{-6}$}
\noindent $= ({0.65\xspace} \pm {0.03\xspace}) \times 10^{-6}$,
and extrapolated to the full kinematic space as
BR$(K^{00}_{\mu 4}) = ({3.45\xspace} \pm {0.10\xspace}_{\rm stat} \pm {0.11\xspace}_{\rm syst} \pm {0.05\xspace}_{\rm ext}) \times 10^{-6}$ = 
$({3.45\xspace} \pm {0.16\xspace}) \times 10^{-6}$, 
using a form factor model based on experimental measurements and ChPT calculations.
The results are consistent with a contribution of the $R$ form factor, as computed at {\em 1-loop} ChPT. The restricted kinematic region considered in this study, $S_\mu > 0.03$~GeV$^2/c^4$, does not allow a more precise measurement of the $R$ form factor description.

\section*{Acknowledgements}
It is a pleasure to express our appreciation to the staff of the CERN laboratory and the technical
staff of the participating laboratories and universities for their efforts in the operation of the
experiment and data processing.

The cost of the experiment and its auxiliary systems was supported by the funding agencies of 
the Collaboration Institutes. We are particularly indebted to: 
the UK Particle Physics and Astronomy Research Council, grant PPA/G/O/1999/00559;
the German Federal Minister for Education and Research (BMBF) under contracts 05HK1UM1/1 and 056SI74;
the Austrian Ministry for Traffic and Research under the contracts GZ 616.360/2-IV and GZ 616.363/2-VIII, and by the Fonds f\"ur Wissenschaft und Forschung FWF Nr. P08929-PHY;
INFN  (Istituto Nazionale di Fisica Nucleare),  Italy;
CERN (European Organization for Nuclear Research), Switzerland; 
ERC (European Research Council)  ``KaonLepton'' starting grant 336581, Europe.

Individuals have received support from:
the Bulgarian National Science Fund under contract DID02-22;
the Royal Society  (grants UF100308, UF0758946), United Kingdom;
ERC Starting Grant 336581, Europe.

\bibliographystyle{new_elsarticle-num} 
\bibliography{kmu400Bib}

\end{document}